\documentclass[12pt]{article}
\usepackage{graphicx}
 \usepackage{longtable}
\usepackage{multirow}
\relax
\def\mod{{\rm~mod~}}

\usepackage{amssymb}
\makeindex
\begin{document}

\hbox{\hskip 12cm NIKHEF/2011-31  \hfil}
\hbox{\hskip 12cm December 2011  \hfil}

\vskip 1.5in

\begin{center}
{\Huge \bf The Kreuzer bi-homomorphism}

\vspace*{.4in}
{\large A.N. Schellekens}
\\
\vskip .2in

{\em Nikhef Theory Group, \\ 
Science Park 105, \\
1098 XG Amsterdam, The Netherlands} \\

\vskip .2in

\vspace*{0.3in}
{\small
This is a brief review of simple current techniques and their application to string theory constructions, centered
around Max Kreuzer's contribution to this field\footnote{Contribution to the Memorial Conference for Max Kreuzer, Vienna, 25-28 June 2011.}.
}

\end{center}

\vskip 1in

\noindent
\newpage

\section{Introduction}

The only paper\cite{Kreuzer:1993tf}
I co-authored with Max Kreuzer is rather atypical for his work, in that it is
algebraic rather than geometrical. In the history of the subject that I am discussing, it plays
an interesting r\^ole. The subject is that of simple currents in  conformal field theory. The name
``simple current" refers to a primary field of a (usually rational) two-dimensional conformal
field theory that has the simplest possible fusion rules with any other field. The name was introduced by
Shimon Yankielowicz and myself in a paper we wrote in 1989 \cite{Schellekens:1989am}. It is a subject that has dominated my 
scientific career since then, even though on several occasions  I wanted to move towards something else.
However, each time someone drew my attention to something else that could still be done in this area.
Max was one of those people. The paper we wrote closed the first stage of research, that of finding
all simple current MIPFs (Modular Invariant Partition Functions), and formed the basis for later developments,
in particular work on boundary and crosscap coefficients for open and unoriented strings. I will use this 
opportunity to review the developments in this area that took place before and after that paper.  

\section{MIPFs and Simple Currents}

In 1989 there was a lot of interest in the problem of finding some or all modular invariant partition
functions of a given two-dimensional conformal field theory (CFT). Modular invariance is a one-loop consistency
condition for closed strings, and at that time the only string theory considered to be of interest was the $E_8 \times E_8$
heterotic string theory, a theory of closed strings. The requirement of modular invariance was initially applied to 
two-dimensional CFTs built out of free bosons and free fermions. For free bosons there is a beautiful and completely
general solution due to Narain \cite{Narain:1985jj} in terms of Lorentzian even self-dual lattices. Although his original work concerned non-chiral theories with $N=4$ supersymmetry, in \cite{LLS} 
it was shown how to use Narain lattices to build chiral heterotic strings. A crucial ingredient in that work was the so-called
bosonic string map, which allows mapping the heterotic string to a bosonic one, where a Narain lattice can be used. In the same year 1986,
two groups \cite{Kawai:1986va} developed a method to satisfy the constraints of modular invariants using free fermions on the world sheet.

But all of this work from 1986 was limited to free two-dimensional CFTs. In 1987 Doron Gepner demonstrated \cite{Gepner:1987vz}
how one could
use interacting CFTs, namely N=2 minimal models, to build string theories. A crucial ingredient in his construction was the
same bosonic string map, this time used to map type-II strings to heterotic ones. This is needed because the only MIPFs
that one can write down without further information are the diagonal one (and the charge conjugation invariant). Hence one
must start with a left-right symmetric CFT, such as a type-II theory. 

An N=2 minimal model is an example of a rational CFT (RCFT). These are conformal field theories with enough additional
world sheet symmetries that the total number of unitary representations is finite. Without additional world-sheet symmetries, {\it i.e.}
with only conformal symmetry, the only such theories are the Virasoro minimal models. Because a fermionic string theory requires
world-sheet supersymmetry, the simplest building blocks one can use to construct them are $N=1$ minimal models. If one wants
to build theories with space-time supersymmetry, one must use $N=2$ building blocks. The simplest ones are the minimal models,
with central charge
\begin{equation}
c=\frac{3k}{k+2} \ \ \ \ \hbox{for} \ k=2 \ldots \infty 
\end{equation}
Since this is always less than the central charge 9 needed to build fermionic strings in four dimensions, one considers
tensor products, whose central is the sum of those of the building blocks. There are 168 combinations of building blocks. 
Using diagonal MIPFs one then gets 168 different four-dimensional heterotic string theories.  

But already at that time it was clear that diagonal MIPFs was not all there was. 
In general, the problem is to find a non-negative integer matrix $M_{ij}$, $i,j=0,\ldots,P-1$ with $ M_{00}=1$, where
$P$ is the number of primary fields. The matrix $M$ must commute with $S$ and $T$, the matrix representation of the modular
transformations $\tau \rightarrow -1/\tau$ and $\tau \rightarrow \tau+1$ on the $P$ Virasoro characters $\chi(\tau)$ of the CFT. In 1987 Cappelli, Itzykson and Zuber\cite{Cappelli:1987xt} classified
all the MIPFs of $SU(2)$ WZW models, which admitted an ADE labelling (see also \cite{Gepner:1987sm}). This naturally raised the question what
possibilities might exist for other WZW models. 
Many papers appeared that found series or isolated examples of MIPFs far various CFTs, see {\it e.g.} 
\cite{Bernard:1986xy,Altschuler:1987gm}.

Another input for our work was a 1988 paper by Erik Verlinde \cite{Verlinde:1988sn}, in which he expressed the coefficients appearing in the BPZ fusion
rules in terms of the modular transformation matrix $S$. The fusion rules may be written formally as
\begin{equation}
[i] \cdot [j] = \sum_k N_{ij}^{~~k}[k]
\end{equation}
which means that the non-negative integer $N_{ij}^{~~k}[k]$ specifies the number of ways the chiral algebra
representations denoted $[i]$ and $[j]$ can couple to the chiral algebra representation $[k]$. Verlinde's result states that this
can be expressed as
\begin{equation}
N_{ij}^{~~k} = \sum_m \frac{S_{im} S_{jm} S_{km}^*}{S_{0m}}
\end{equation}

We realized that something special happens if the fusion rules are as simple as they can be. In general,  the fusion of any two
representations must yield {\it at least} one other representation. 
There is {\it precisely} one if either $[i]$, $[j]$ or $[k]$ are equal to the identity,
$[0]$, but often there are special representations $[J]$ that have the property that
\begin{equation}
[J] \cdot [j] = [Jj] \ \ \ \hbox{for all} \ j, 
\end{equation}
where ``$Jj$" is merely a convenient notation for the unique fusion product of $J$ and $j$. We called such representations (or the
corresponding primary fields) ``simple currents". The reason for the adjective is obvious, and the name ``current" will be explained 
below. If a primary field $[J]$ has this property, this has an interesting consequence for its operator products. Since in any
operator product  with a field $[j]$ only  $[Jj]$ and its descendants can appear, there is a well-defined monodromy phase when
transporting $[J]$
on a contour around $[j]$  in the complex plane. On other words, the OPE has a well-defined branch cut, unlike a generic OPE.
Note that all these manipulations are performed in a chiral half of the theory. 
This allows us to define a {\it monodromy charge}
\begin{equation} 
Q_J(i) = h_i + h_J - h_{Ji}
\end{equation}
This charge is conserved by all correlation functions, and hence defines a symmetry. The final step is then to
perform an orbifold-like procedure and ``mod out" that symmetry. In general this leads to a new theory with a new MIPF.
If $h_J \in \mathbb{Z}$ this new MIPF defines an extension of the chiral algebra by the chiral field with label $[J]$, and one normally
refers to such a field as a ``current" in the extended algebra. Hence our name ``simple current". If $h_J$ is fractional this name
is perhaps less suitable, although it is customary to talk about fermionic currents, supercurrents and parafermionic currents.
After our paper had appeared, we learned that Fuchs and Gepner had already observed  the appearance of fields with simple
fusion in a study of four-point functions \cite{Fuchs:1986ey}.

In a subsequent paper \cite{Schellekens:1989dq} we wrote the corresponding symmetry as a relation between matrix elements of $S$, and used that to
check the conjectured modular invariance explicitly. This symmetry was discovered independently by Kenneth Intriligator 
\cite{Intriligator:1989zw}, who
called it ``bonus symmetry". It relates elements of $S_{ij}$ on simple current orbits
\begin{equation}\label{Ssym}
S_{Ki,j}=e^{2\pi Q_K(j)} S_{ij}
\end{equation}
In \cite{Schellekens:1989dq} we also wrote down an explicit formula for the matrix $M$:
\begin{eqnarray*}\label{MatrixM}
M_{J^pi,J^qj}=\delta_{ij}\sum_{\ell=1}^{N}\delta^{N_j}_{q,p+\ell} \delta^1\left[Q_J(i)+
\left(\frac{2p+\ell}{2N}\right)r\right]
\end{eqnarray*}
where the ``monodromy parameter" $r$ is defined by
\begin{eqnarray*}\label{diagmod}
h_J = \frac{r (N-1)}{2N} \ \rm{mod}\ 1
\end{eqnarray*}
Note that $r$ is defined modulo $2N$ if $N$ is even.
This formalism includes essentially all prior work on MIPFs as special cases, with the exception of the 
exceptional invariants. Strictly speaking this is a tautology, because it is precisely the definition 
of exceptional invariants, but the point is that in all cases studied thus far the exceptional invariants 
are rare. In the example of the ADE classification of $SU(2)$ only the three ``E" invariants are exceptional. 

The power of the simple current formalism lies in the fact that it can be applied to any CFT with simple currents,
without detailed knowledge of the matrix $S$. In reality, the vast majority of the RCFT's we know are WZW-model
based (for example coset CFTs), and one might argue that in those cases we already know $S$ explicitly anyway.
However it is still a great convenience to be able to write down a huge number of MIPFs without invoking a formula for $S$.
Nowhere is this demonstrated more clearly that in the application to Gepner models, which Shimon Yankielowicz and I attacked
later in 1989 \cite{Schellekens:1989wx}. Gepner models have huge simple current groups: each minimal model factor has $4k+8$ simple currents.
Not only were we able to vastly extend the total number of MIPFs, but this also lead to a significant simplification
 of the formalism and to new insights. The generalized GSO projection needed to get space-time supersymmetry was realized as a 
 simple current extension of the chiral algebra, and it became obvious that the simple current formalism allowed us to make
 this extension only in the fermionic sector, and use a higher spin current (not appearing in the massless spectrum) in the bosonic
 sector. In this way a large number of spectra with $SO(10)$ unified gauge symmetry -- instead of the less desirable $E_6$
 gauge symmetry occurring in (2,2) models -- could be constructed. We called these ``$(1,2)$" models, where the ``1" indicates that
 world-sheet in the left-moving, bosonic sector remained unbroken. We produced a huge ``phonebook" of such spectra, which is
 available in pdf form (see \cite{T58}), but unfortunately not in a computer readable format. For (2,2) spectra similar lists were presented in
  \cite{Fuchs:1989yv}. Earlier results on subsets  of MIPFs were presented in \cite{Lutken:1988hc} and \cite{Lynker:1988fs}. 
 
 The method for building simple current MIPF used in \cite{Schellekens:1989wx} was still a bit primitive, though. We used the simple current matrix
 $M(J)$ defined in Eqn. (\ref{MatrixM}) above, but we used the fact  that we could also multiply such matrices. So we considered products
 \begin{eqnarray*}
P(\tau,\bar\tau)= \chi(\tau) M(J_1)  M(J_2) \ldots  M(J_k) \bar\chi(\bar\tau)
\end{eqnarray*}
combined with a normalization condition for the identity. It was not clear how many such matrices one could multiply 
and still get something new, nor was it clear if anything was missed this way. 

In 1990 it became clear that indeed something {\it was} missed this way. Up to then, integer spin simple currents were
believed to yield extensions of the chiral algebra. Multiplying such MIPFs then only gives extensions as well. But it turned
out that in certain cases automorphism MIPFs were also possible. An automorphism MIPF is characterized by a matrix $M$ that
acts as a permutation of the characters. A year later Beatriz Gato-Rivera and I started thinking about the classification problem
of all simple current MIPFs \cite{GatoRivera:1991ru}. We tried classifying all MIPFs with matrices $M_{ij}$ that had vanishing entries if $i$ and $j$ are
not linked by simple currents. We succeeded in part, achieving a full classification for simple current groups
$(\mathbb{Z}_p)^k$, with $p$ prime (the result is also valid if $p$ is a product of prime factors, each occurring only once, since
the entire problem then just factorizes, but it is not valid if there are factors $\mathbb{Z}_{p^n}$, $n > 1$). We derived a formula for the
total number of MIPFs
\begin{equation}\label{Total}
T(\rho,k,p)=\prod_{\ell=0}^k(1+p^{\ell})
\end{equation}
where $\rho$ is the monodromy matrix of the currents
\begin{eqnarray*}
Q_i(J_j)=Q_j(J_i)\equiv R_{ij} =\frac{1}{p} \rho_{ij}\ \ , \rho_{ij} \in \mathbb{Z}
\end{eqnarray*}
Note that $\rho_{ij}$ is defined modulo $N_i$ as well as $N_j$. If $N_i$ is even, there is an additional requirement that
the diagonal elements $\rho_{ii}$ must be equal to $r_i$ , the monodromy parameter introduced in (\ref{diagmod}); this may
require shifting $\rho_{ii}$ by an amount $N_i$.  The result (\ref{Total}) holds if all $\rho_{ii}$ are even. Since this can always
be achieved for $N_i$ odd by shifting by $N_i$, this condition matters only for $N_i$ even.
Note that the total number of MIPFs does not depend on $\rho$ at all. This was merely an observation, but it was not 
understood.

This is where Max entered the scene. It is now so long ago that I do not even remember the reason I met Max. Presumably he
had been invited to NIKHEF to give a seminar. I also do not recall much of the work itself, except that it started with Max's
suggestion to add discrete torsion to the orbifold procedure used in derive the simple current MIPF. This led to two important
benefits: first of all we were able to extend the classification to all simple current groups, not just $(\mathbb{Z}_p)^k$, and
secondly monodromy independence followed straightforwardly, because the orbifold procedure did not care about monodromies,
except for the even integer subtleties mentioned above. This led to a general and complete formalism for simple current MIPFs which
is still used today. Because it plays a pivotal r\^ole I quote the main result in its orginal form:

Suppose one has a conformal field theory with simple currents
generating a center ${\cal C}$.
Then the complete set of simple current invariants
of that theory can be obtained by the following procedure
\begin{itemize}
\item{Choose any subgroup ${\cal H}$ of ${\cal C}$}
\item{Choose a basis of currents $J_1, \ldots, J_k$ that generate ${\cal H}$.}
\item{Compute the current-current monodromies $R_{ij}$ in that basis.}
\item{Choose any properly quantized matrix $X$ whose
symmetric part is $\frac12 R \mod 1$ (in other words $X+X^T = R $).}
\end{itemize}
The modular invariant partition function corresponding to this choice
is then given by a matrix whose only non-zero elements are
\begin{equation}\label{GenInv}
 M_{a,[\vec \beta]a} = {\rm Mult}(a) \prod_i \delta^1 ( Q_i (a) + X_{ij} \beta_j )\ ,
\end{equation}
where $\delta^1$ is equal to 1 if its argument is an integer, and
vanishes otherwise. The factor ${\rm Mult}(a)$ appears because $a$ may be
a fixed point of some currents.
In that case the $\beta$-sum in (\ref{GenInv}) includes all
terms involving $a$ more than once, and ${\rm Mult}(a)$ is the number of times
this happens.  This is the generalization of (\ref{MatrixM}) to more than one
factor.

This paragraph was lifted directly from page 17 of the  original paper. By ``center" we meant the complete discrete group of simple currents;
$[\vec \beta]a$ stands for $J_1^{\beta_1} \ldots J_k^{\beta_k}a$.
 Some subtleties regarding even integers can be found on the same page 17 in the original paper, but will not be repeated here.
 
 The definition of ``properly quantized" is that $X$ must be a matrix of rational numbers satisfying $N_i X_{ij} \in \mathbb{Z}$,  
 $X_{ij}  N_j\in \mathbb{Z}$ (no summation implied). This matrix $X$ is the object appearing in the title of this talk. Many years later,
 the authors of \cite{Fuchs:2004dz} defined
 an object they called $\Xi$, related to $X$ as 
 \begin{equation}
 \Xi(g,h)={\rm exp} \left( 2\pi \sum_{a,b=1}^k m_a X_{ab} n_b \right)
 \end{equation}
 where $g$ and $h$ are elements of the simple current group written in the form $g=\prod_a (g_a)^{m_a}$ and $g=\prod_a (h_a)^{n_a}$.
 Obviously $\Xi(g,h)$ defines a bi-homomorphism on the simple current group, which the authors referred to as KSB, where
 ``K" stands for Kreuzer and ``B" for bi-homomorphism. We will see this quantity return later in this paper.
 
 Recently we have applied this formalism to heterotic strings \cite{GatoRivera:2010gv}, in order to go a few steps beyond what was possible in 1989. In the
 course of doing that we re-obtained the results obtained in 1989 using matrix products. The results of 1989 were obtained by a deep,
 but randomized search, which was basically continued until the saturation point appeared to be reached
 (no new spectra appeared during some fixed time).  
 We have not checked the entire list systematically
 (since it is not available electronically anymore this would be difficult to do), but very few cases, if any, are missing. This
 implies that all MIPFs that in principle cannot be obtained using matrix multiplication, but only using the complete formalism,
 can apparently still be obtained in different ways. This must be due to peculiar degeneracies in the Gepner models, which are
 only partly understood. 
 
 In the new complete scan of all $(2,2)$  symmetric as well as asymmetric MIPFs we found a total of 906 distinct Hodge number pairs. If we also consider 
the number of singlets and massless vector bosons the total number of distinct $(2,2)$ spectra is 2013. The number of Hodge pairs can
be compared with similar numbers for free fermionic $(2,2)$ models,  and numbers from Max Kreuzer's homepage \cite{MK}. For free fermions
we obtained 26 distinct Hodge pairs in \cite{Kiritsis:2008mu} (see also \cite{Donagi:2008xy,Ploger:2007iq} on related  
$\mathbb{Z}_2 \times \mathbb{Z}_2$ spectra). Kreuzer and Skarke \cite{Kreuzer:2000xy} obtained 30108
 distinct Hodge pairs in their work on 
reflexive polyhedra. Furthermore a list of 4370 Hodge number pairs are listed in \cite{MK} for Landau-Ginsburg models
(after combining the ``Untwisted", ``Abelian" and ``Discrete torsion" databases). This clearly illustrates the naive expectation that
free theories are too special, that the geometric approach is more general, and the interacting CFTs are somewhere in between:
almost the geometric mean, in fact. As might be expected, the list of 4370 LG Hodge number pairs includes all 906 Gepner Hodge number 
pairs, which in its turn includes the 26 free fermion Hodge number pairs. 

\section{Fixed point resolution}

An important element in the story, although not directly related to Max's contribution, is fixed point resolution. 
This problem arises if the mul\-tipli\-ci\-ties ${\rm Mult}(a)$ in the main formula above are larger than 1, which in its turn is
a consequence of simple currents having fixed points: $Ja=a$. In MIPFs this merely leads to integer mul\-tipli\-ci\-ties larger than 1,
but these multiplicities require further attention if one attempts to write down the modular transformations for the characters of
the new, extended CFT. They lead to degeneracies: several characters of the new CFT are equal to one and the same character
of the original CFT. 
 Therefore modular transformations of the original CFT do not provide sufficient information to determine those of
 the extended CFT. 
 
 Based on early conjectures in a paper \cite{Schellekens:1989uf} with Shimon Yankielowicz I wrote in 1989, J\"urgen Fuchs, Christoph Schweigert and I 
 \cite{Fuchs:1996dd} were
 able to solve this problem in the middle of the nineties, for WZW based theories. The missing information is contained in modular transformation matrices
 $S^J$ of a different algebraic object, which is often, but not always, itself a CFT. It is obtained as follows. The simple current $J$
 of a WZW model defines (with one exception, $E_8$ level 2) an automorphism of the extended Dynkin diagram underlying the WZW model.
 One can fold up the Dynkin diagram according to this automorhism, according to specific rules spelled out in 
 \cite{Fuchs:1995zr}. This yields the extended Dynkin diagram of the ``algebraic object" alluded to above. Usually this is an affine
 Lie algebra, but in some cases (the ones that were problematic in \cite{Schellekens:1989uf}) it is a twisted affine algebra, which is not a CFT, but still
 has the remarkable property that it has an associated modular group representation that can be used to define $S^J$. 
 
 The main result of \cite{Fuchs:1996dd}  is as follows. Characters of the extended theory are labelled by some representation label
 $a$ of the original CFT, plus a degeneracy label $i$ distinguishing the individual representations of the extended CFT related to
 the same character in the original CFT. The modular transformation matrix $\tilde S$ of the extended CFT is given by:
\def\U{{\cal U}}
\def\S{{\cal S}}
\def\G{{\cal H}}
\begin{equation}
\tilde S_{(a,i),(b,j)}=
  {|\G| \over \sqrt{|\U_a|\,|\S_a|\,|\U_b|\,|\S_b|}}\,
  \sum_{J \in \G} \Psi^a_i(J)\, S^J_{a,b}\, \Psi^b_j(J)^*\
\end{equation}
Here $\G$ is the simple current subgroup used to define the MIPF, $\S_a$ is the stabilizer of $a$ (the set of simple
currents $J$ with $J_a=a$), and $\U_a$ is a subgroup of $\S_a$, called the ``untwisted stabilizer", 
and the $\Psi^a$'s are characters of the group $\U_a$.  The untwisted stabilizer is the subgroup on which a computable
quantity called the twist $F(a,K,J)$ (which can be $+1$ or $-1$) is equal to 1. This twist is a factor that modifies 
(\ref{Ssym}) for fixed point resolution matrices $S^J$:
\begin{equation}
S^J_{Ki,j}=F_i(K,J) e^{2\pi i Q_K(j)}S^J_{ij}
\end{equation}
and then the formal definition of the untwisted stabilizer is
\begin{equation}
{\cal U}_i =\left\{ J \in {\cal S}_i \| F_i(K,J) = 1,\ \hbox{for all}\ K \in {\cal S}_i\right\}
\end{equation}

This formula holds for all simple current extensions of the chiral algebra. Since these only involve mutually local integer spin currents,
the general simple current formalism of \cite{Kreuzer:1993tf} is not needed to obtain them. However, we will see that all 
quantities introduced here will make a re-appearance later.

\section{Boundary Conformal Field Theory}

All previous results were for conformal field theory on closed surfaces. Work by the
Tor Vergata group in Rome in the early nineties pointed out an interesting application to
CFT on surfaces with boundaries and crosscaps. I am grateful to Augusto Sagnotti for bringing
this to my attention, and urging me to work on it.

In 1989, Cardy had written a breakthrough paper on CFT boundary conditions in an attempt to
understand the Verlinde formula. The Rome group started applying that work to other cases, and
to non-orientable surfaces. The ``other cases" usually were simple current MIPFs, in particular those
of the $SU(2)$ WZW model and the discrete series. They pointed out an interesting connection with
fixed point resolution in \cite{Bianchi:1991rd}. Intuitively, the reason this matters is that if a
field scatters of a boundary, one has to be able to decide into which of several degenerate fields it is
reflected. It turns out that this choice is governed by the same fixed point resolution matrices $S^J$ 
introduced above. 

After the Rome group had worked out several series of examples, the natural question arose how
this would work in general. Or more precisely: given a CFT and a simple current MIPF of the most
general form, as given by the KS formalism, what is the set of boundary and crosscap coefficients
that go with it. 

This question bifurcated into several others. The first one was to determine which closed string states
actually propagate in the closed channel between two boundaries (the transverse channel of the annulus).
These states are called ``Ishibashi states" \cite{Ishibashi:1988kg}.
In Cardy's work they are in one-to-one correspondence with the set of primary fields. This happens because he
used the charge conjugation modular invariant, with $M_{mm^c}=1$,  and all other elements of $M_{mn}$ vanishing
($m^c$ is the charge conjugate of $m$). A plausible guess is that in non-trivial MIPFs  the number of Ishibashi states
for a given label $i$ is equal to the value of $M_{mm^c}$, which now can in principle be any non-negative integer. This
is indeed correct. Hence this number is given by the factor ``Mult($a$)" above. This implies that the Ishibashi states can be labelled
by elements of the Stabilizer ${\cal S}_m$ of $m$. So we may introduce a set of labels $(m,J)$, $J \in {\cal S}_m$, with
\begin{equation}
Q_L(m) + X(L,J) = 0 \mod 1\ \hbox{for all} \ L \in {\cal H}
\end{equation}
here ${\cal H}$ is the simple current subgroup that defines the MIPF, and $X$ is the rational matrix related 
to the KSB. The MIPF is implicitly assumed to be multiplied by a charge conjugation MIPF.  

The next, far less trivial question is which boundaries can be used. What we are looking for is the complete set
of boundary states that respect all the symmetries of the original CFT, without taking into account the MIPF, even if
the MIPF itself  implies a chiral algebra extension. In a paper from 1996 \cite{Pradisi:1996yd}, Pradisi, Sagnotti and Stanev
had introduced the important concept of completeness of boundaries. This states that the set of boundary states forms
a complete basis in the space of transverse channel states, and hence there should be as many boundary states as there
are Ishibashi states. In simple examples one could fulfill that condition by associating one boundary label with every simple
current ${\cal H}$-orbit (including orbits that are projected out in the MIPF). But is was far less obvious how to deal with fixed points.
After a lot of experimentation, mainly done in several papers of the Rome group, and by J\"urgen Fuchs and Christoph Schweigert,
the following description was arrived at in \cite{Fuchs:2000cm}. The boundary states can be labelled by elements of a discrete
group we called the ``central stabilizer", which is defined as follows
\begin{eqnarray*}
{\cal C}_i &=&\left\{ J \in {\cal S}_i \| F^X_i(K,J) = 1,\ \hbox{for all}\ K \in {\cal S}_i\right\}\\
F^X_i(K,J)&=&e^{2\pi i X(K,J)} F_i(K,J)^*\\
\end{eqnarray*}
Here the KSB makes its appearance.

Note that the rules defining Ishibashi states look rather different from those defining boundary states, and that it is
far from obvious that the two set have the same size. This can be proved by writing down the boundary coefficients that
relate the two basis, and showing that this matrix is invertible. The result is
\begin{equation}
B_{[a,\psi_a](m,J)}=\sqrt{\frac{|{\cal H}|}{{|\cal C}_a| |{\cal S}_a|}}
\psi^*_a(J)S^J_{am}
\end{equation}
Here $\psi_a(J)$ is a character of the discrete group ${\cal C}_a$, which turns out to be the most convenient way to
label the boundary states. Apart from invertibility there is a much more restrictive test the coefficients have to satisfy
namely integrality and non-negativity of the annulus coefficients
\begin{equation}
A^{i}_{~ab}=\sum_m \frac{S_{im} B_{am}B_{bm}}{S_{0m}}  \ ,
\end{equation}
where $a$ and $b$ are a short-hand notation for the labels $[a,\psi_a]$. All this was checked explicitly in the thesis of my
student Lennaert Huiszoon \cite{Huiszoon}.

The next question to be answered was: which orientifolds are allowed for a given KS simple current MIPFs, and what is
the expression for the crosscap  coefficients that describe the reflection of Ishibashi states from a crosscap (a hole in a Riemann
surfaces with opposite sides identified in an orientation-reversing way). As usual, the first answer was  given by the Rome group,
who showed that for the charge conjugation invariant it could be expressed in terms of the matrix element $P_{i0}$ of the
``P-matrix" $P=\sqrt{T}ST^2S\sqrt{T}$. They also found examples where several distinct orientation reversals were allowed, and 
worked out the result for D-invariants of $SU(2)$. Starting from there, with my students Huiszoon and Sousa we managed to
derive the result for more general simple current MIPFs, and the final form was written down in \cite{Fuchs:2000cm}.
\begin{equation}
\Gamma_{(m,J)}=\frac{1}{\sqrt{|{\cal H}|}} \sum_{L \in {\cal H}} \eta(K,L)
P_{LK,m} \delta_{J,0}
\end{equation}
where 
\begin{equation}
\eta(K,L)=e^{i\pi(h_K-h_{KL})}\beta_K(L)
\end{equation}
\begin{equation}
\beta_K(J)\beta_K(J')=\beta_K(JJ') e^{2\pi i X(J,J')}; J , J' \in {\cal H} \ , 
\end{equation}
where once more we see the KS bi-homorphism appear. The simple current $K$ is called
the ``Klein bottle current", and the equation for the coefficients $\beta_K$ (which can be signs or $\pm i$)
may allow several solutions differing by signs. There are some further conditions, and some of the choices are
equivalent, see section 2 of \cite{Dijkstra:2004cc} for further details. These crosscap coefficients appear in the
formulas for the Klein bottle and Moebius multiplicities, which are subject to severe integrality constraints, 
checked explicitly in \cite{Huiszoon} 
\begin{eqnarray*}
K^i&=&\sum_m \frac{S_{im} \Gamma_m\Gamma_{m}}{S_{0m}}\\ 
M^{i}_{~a}&=&\sum_m \frac{P_{im} B_{am}\Gamma_{m}}{S_{0m}}\\
\end{eqnarray*}

\section{Applications}

The formalism explained above can be put to practical use in the construction of 
exact perturbative string spectra in an area we still no little about: interacting two-dimensional CFT's. 
The vast majority of exact results has been obtained for free fermions or (twisted) free bosons (orbifolds).  
On the other side of the range of possibilities are geometric methods, such as Calabi-Yau manifold compactifications
or F-theory. These cover undoubtely a larger piece of the landscape, but exact results can only be obtained for a limited
number of quantities. Interacting CFTs provide a middle ground between the two. Because of the naive notion that these
provide ``rational points in continuous moduli spaces" one would have expected this to be a extremely large set, but in reality
the class of accessible rational CFTs is rather limited. By ``accessible" I mean here that the full set of modular data is available:
the exact ground state weights and dimensions and the matrices $S$, $T$ and $S^J$. Until recently, this was true only for
tensor products of $N=2$ minimal models. Recently, with my student Michele Maio \cite{Maio:2010eu}, we succeeded in adding one more 
building block to this list, the $\mathbb{Z}_2$ permutation orbifolds of two identical, already known building blocks. 

We have explored the RCFT orientifold and heterotic landscape in recent work \cite{Dijkstra:2004cc, GatoRivera:2010gv}. 
The general simple current formalism 
described in \cite{Kreuzer:1993tf}, and its open string analog \cite{Fuchs:2000cm}
 allowed us to survey a much larger part of the landscape than would otherwise be possible.   The features we look for, chiral
 spectra containing an $SU(3)\times SU(2)\times U(1)$ gauge group and three families are not rare in the string theory landscape,
 but they are still statistically challenged by a factor about  $10^{-5} \ldots 10^{-15}$ depending on the starting assumptions. 
 Without this formalism, even scratching off this small layer from the surface would not have been possible.
 
 Almost twenty years after \cite{Kreuzer:1993tf} was written, Max lives on through his work.

 \section*{Ackowledgements}
 
 I would like to thank the organizers of this meeting for giving me the opportunity to
 present this work and pay tribute to Max Kreuzer's important contribution to it. 
 
\bibliographystyle{ws-rv-van}
\bibliography{ws-rv-sample}

\begin{thebibliography}{99}

\bibitem{Kreuzer:1993tf}
  M.~Kreuzer and A.~N.~Schellekens,
   Nucl.\ Phys.\  B {\bf 411} (1994) 97

\bibitem{Narain:1985jj}
  K.~S.~Narain,
  Phys.\ Lett.\  B {\bf 169} (1986) 41.


\bibitem{LLS}
W.~Lerche, D.~L\"ust and A.~N.~Schellekens,
  Nucl.\ Phys.\ B {\bf 287} (1987) 477. 

\bibitem{Kawai:1986va}
  H.~Kawai, D.~C.~Lewellen and S.~H.~H.~Tye;\\
  I.~Antoniadis, C.~P.~Bachas and C.~Kounnas,
  Nucl.\ Phys.\  B {\bf 289}, 87 (1987).



\bibitem{Gepner:1987vz}
  D.~Gepner,
  Phys.\ Lett.\  B {\bf 199} (1987) 380.

\bibitem{Cappelli:1987xt}
  A.~Cappelli, C.~Itzykson, J.~Zuber,
  Commun.\ Math.\ Phys.\  {\bf 113 } (1987)  1.
  
\bibitem{Bernard:1986xy}
  D.~Bernard,
  Nucl.\ Phys.\  {\bf B288 } (1987)  628.
  
\bibitem{Altschuler:1987gm}
  D.~Altschuler, J.~Lacki, P.~Zaugg,
  Phys.\ Lett.\  {\bf B205 } (1988)  281.
  
  
\bibitem{Gepner:1987sm}
  D.~Gepner,
  Nucl.\ Phys.\  {\bf B290 } (1987)  10.

\bibitem{Verlinde:1988sn}
  E.~P.~Verlinde,
  Nucl.\ Phys.\  {\bf B300 } (1988)  360.


\bibitem{Schellekens:1989am}
  A.~N.~Schellekens and S.~Yankielowicz,
  Nucl.\ Phys.\  B {\bf 327} (1989) 673.

\bibitem{Schellekens:1989dq}
  A.~N.~Schellekens and S.~Yankielowicz,
  Phys.\ Lett.\  B {\bf 227} (1989) 387.



\bibitem{Intriligator:1989zw}
  K.~A.~Intriligator,
  Nucl.\ Phys.\  B {\bf 332} (1990) 541.
  
\bibitem{Fuchs:1986ey}
  J.~Fuchs, D.~Gepner,
  Nucl.\ Phys.\  {\bf B294 } (1987)  30.

  
\bibitem{Schellekens:1989wx}
  A.~N.~Schellekens and S.~Yankielowicz,
  Nucl.\ Phys.\  B {\bf 330} (1990) 103.

\bibitem{Fuchs:1989yv}
  J.~Fuchs, A.~Klemm, C.~Scheich, M.~Schmidt,
  Ann. Phys.\  {\bf 204 } (1990)  1-51.
  
\bibitem{Lutken:1988hc}
  C.~A.~Lutken, G.~G.~Ross,
  Phys.\ Lett.\  {\bf B213 } (1988)  152.
 
  
\bibitem{Lynker:1988fs}
  M.~Lynker, R.~Schimmrigk,
  Phys.\ Lett.\  {\bf B215 } (1988)  681.
 

\bibitem{GatoRivera:1991ru}
  B.~Gato-Rivera and A.~N.~Schellekens,
  Comm.\ Math.\ Phys.\  {\bf 145} (1992) 85.

\bibitem{Fuchs:2004dz}
  J.~Fuchs, I.~Runkel, C.~Schweigert,
  Nucl.\ Phys.\  {\bf B694 } (2004)  277-353.

   \bibitem{T58}
http://www.nikhef.nl/$\sim$t58

\bibitem{GatoRivera:2010gv}
  B.~Gato-Rivera and A.~N.~Schellekens,
  Nucl.\ Phys.\  B {\bf 841} (2010) 100


\bibitem{Kiritsis:2008mu}
  E.~Kiritsis, M.~Lennek, B.~Schellekens,
  JHEP {\bf 0902 } (2009)  030.

\bibitem{Donagi:2008xy}
  R.~Donagi, K.~Wendland,
  J.\ Geom.\ Phys.\  {\bf 59 } (2009)  942-968.

  \bibitem{Ploger:2007iq}
 F.~Ploger, S.~Ramos-Sanchez, M.~Ratz and P.~K.~S.~Vaudrevange,
  JHEP {\bf 0704} (2007) 063
  
   \bibitem{MK}
http://hep.itp.tuwien.ac.at/$\sim$kreuzer/CY/

\bibitem{Kreuzer:2000xy}
  M.~Kreuzer, H.~Skarke,
  Adv.\ Theor.\ Math.\ Phys.\  {\bf 4 } (2002)  1209-1230.

\bibitem{Schellekens:1989uf}
  A.~N.~Schellekens, S.~Yankielowicz,
  Nucl.\ Phys.\  {\bf B334 } (1990)  67.
  

\bibitem{Fuchs:1996dd}
  J.~Fuchs, A.~N.~Schellekens, C.~Schweigert,
  Nucl.\ Phys.\  {\bf B473 } (1996)  323-366.


\bibitem{Fuchs:1995zr}
  J.~Fuchs, B.~Schellekens, C.~Schweigert,
  Comm.\ Math.\ Phys.\  {\bf 180 } (1996)  39-98.

\bibitem{Cardy:1989ir}
  J.~L.~Cardy,
  Nucl.\ Phys.\  {\bf B324 } (1989)  581.

\bibitem{Ishibashi:1988kg}
  N.~Ishibashi,
  Mod.\ Phys.\ Lett.\  {\bf A4 } (1989)  251.


\bibitem{Bianchi:1991rd}
  M.~Bianchi, G.~Pradisi, A.~Sagnotti,
  Phys.\ Lett.\  {\bf B273 } (1991)  389-398.
  
\bibitem{Pradisi:1996yd}
  G.~Pradisi, A.~Sagnotti, Y.~.S.~Stanev,
  Phys.\ Lett.\  {\bf B381 } (1996)  97-104.
  
\bibitem{Fuchs:2000cm}
  J.~Fuchs, L.~R.~Huiszoon, A.~N.~Schellekens, C.~Schweigert, J.~Walcher,
  Phys.\ Lett.\  {\bf B495 } (2000)  427-434.

\bibitem{Huiszoon}
  L.~R.~Huiszoon,
  ``D-branes and O-planes in string theory : an algebraic approach'', PhD thesis (2002)

\bibitem{Dijkstra:2004cc}
  T.~P.~T.~Dijkstra, L.~R.~Huiszoon, A.~N.~Schellekens,
  Nucl.\ Phys.\  {\bf B710 } (2005)  3-57.

\bibitem{Maio:2010eu}
  M.~Maio, A.~N.~Schellekens,
  Nucl.\ Phys.\  {\bf B845 } (2011)  212-245.

\bibitem{GatoRivera:2010gv}
  B.~Gato-Rivera, A.~N.~Schellekens,
  Nucl.\ Phys.\  {\bf B841 } (2010)  100-129; Nucl.\ Phys.\  {\bf B846 } (2011)  429-468;  Nucl.\ Phys.\  {\bf B847 } (2011)  532-548.\\
   M.~Maio, A.~N.~Schellekens,
  Nucl.\ Phys.\  {\bf B848 } (2011)  594-628.


\end{thebibliography}

\end{document}